\documentstyle[twocolumn,aps,epsfig]{revtex}
\begin{document}
\preprint{\vbox{\hbox{IFT--P.021/98}      \vspace{-0.3cm}
                \hbox{IFIC/98--35}        \vspace{-0.3cm}
                \hbox{FTUV/98--34}        \vspace{-0.3cm}
                \hbox{IFUSP/98--1306}   \vspace{-0.3cm} }}

\twocolumn[\hsize\textwidth\columnwidth\hsize\csname @twocolumnfalse\endcsname

\title{Strongly Interacting Vector Bosons at the LHC: \\
       Quartic Anomalous Couplings}

\author{A.\ S.\ Belyaev $^{1,2}$, 
        O.\ J.\ P.\ \'Eboli $^1$, 
        M.\ C.\ Gonzalez--Garcia $^3$, \\ 
        J.\ K.\ Mizukoshi $^4$, 
        S.\ F.\ Novaes $^{1}$, and 
        I.\ Zacharov $^5$ }

\address{$^1$ Instituto de F\'{\i}sica Te\'orica, 
              Universidade Estadual Paulista, \\ 
              Rua Pamplona 145, 01405--900 S\~ao Paulo, Brazil.\\
         $^2$ Skobeltsyn Institute of Nuclear Physics, 
              Moscow State University \\
              119899 Moscow, Russian Federation. \\
         $^3$ Instituto de F\'{\i}sica Corpuscular IFIC/CSIC,
              Departament de F\'{\i}sica Te\`orica \\
              Universitat de Val\`encia, 46100 Burjassot,
              Val\`encia, Spain. \\
         $^4$ Instituto de F\'{\i}sica, Universidade de S\~ao Paulo \\
              C.P.\ 66318, 05315--970, S\~ao Paulo, Brazil. \\
         $^5$ Silicon Graphics European Headquarters, \\
              Chemin des Avouillons 30, 1196 Gland,  Switzerland. } 

\date{May 4, 1998}

\maketitle

\newpage

\begin{abstract}
We analyze the potential of the CERN Large Hadron Collider to
study anomalous quartic vector--boson interactions through the
production of vector--boson pairs accompanied by jets. In the
framework of $SU(2)_L \otimes U(1)_Y$ chiral Lagrangians, we
examine all effective operators of order $p^4$ that lead to new
four--gauge--boson interactions but do not alter trilinear
vertices. In our analyses, we perform the full tree level
calculation of the processes leading to two jets plus
vector--boson pairs, $W^+W^-$, $W^\pm W^\pm$, $W^\pm Z$, or $ZZ$,
taking properly into account the interference between the
standard model and the anomalous contributions. We obtain the
bounds that can be placed on the anomalous quartic interactions
and we study the strategies to distinguish the possible new
couplings.
\end{abstract}

\vskip2pc]
\narrowtext

\section{Introduction}

The standard model (SM) of the electroweak interactions, based on
the $SU(2)_L \otimes U(1)_Y$ gauge symmetry, has accomplished an
impressive agreement of its predictions for the fermion--vector
boson couplings with all the recent experimental data
\cite{ewwg}. Notwithstanding, the tests of the triple and quartic
bosonic interactions still lack the same accuracy to further
confirm the local gauge invariance of the theory or to indicate
the existence of new physics beyond the SM.

The interactions responsible for the electroweak symmetry
breaking play an important r\^ole in the gauge--boson scattering
at high energies as they are an essential ingredient to avoid
unitarity violation in the scattering amplitudes of massive
vector bosons at the TeV scale \cite{unit}. There are two
possible forms of electroweak symmetry breaking which lead to
different solutions to the unitarity problem: $(a)$ there is a
scalar particle lighter than 1 TeV, the standard model Higgs
boson, or $(b)$ such particle is absent and the longitudinal
components of the $W$ and $Z$ bosons become strongly interacting
at high energies. In the latter case, the symmetry breaking
occurs due to the nonzero vacuum expectation value of some
composite operators which are related with new underlying
physics.

In this work we analyze the potential of the CERN Large Hadron
Collider (LHC) to study deviations of the quartic vector--boson
couplings from the SM predictions, assuming a strongly
interacting electroweak symmetry breaking sector (SEWS). In fact,
the LHC will be the first collider capable of directly studying
these couplings through the scattering of gauge--bosons in
reactions like $pp \rightarrow q q V V \rightarrow V V j j$
\cite{bagger0,bagger,dobado}, with $V = W^\pm$ or $Z^0$. Studies
of quartic couplings will also be possible at future $e^+e^-$
colliders \cite{lih,barger,boos,he,vvv}, and also in $e\gamma$
\cite{e:g} and $\gamma\gamma$ collisions \cite{g:g}.
Notwithstanding, at present, this sector of the SM can only be
indirectly bounded by the precise measurements of the electroweak
parameters \cite{loops,he:yuan}.

In this paper we assume that there are no new heavy resonances
at the LHC energy scale, which means that the $SU(2)_L \otimes
U(1)_Y $ gauge symmetry is nonlinearly realized. In this case,
the electroweak sector must be parametrized in terms of
electroweak chiral Lagrangians. We study the complete set of
dimension four operators contributing only to quartic
vector--boson couplings and we estimate the sensitivity of the
LHC to search for deviations from the SM predictions.

We present the results for the full tree level calculation of the
processes $pp \to V V +$ 2 jets, with $V = W^\pm, Z^0$, taking
properly into account the interference between the SM and
anomalous quartic contributions. This improves the previous
studies of SEWS at the LHC \cite{bagger0,bagger,dobado} which
relied upon the equivalence theorem \cite{equ} or/and the
effective $W$--boson approximation \cite{evb}.  Moreover, we
performed our calculation both in the unitary and 't
Hooft--Feynman gauges, and we also included the efficiencies for
detecting the leptons originating from the vector boson decays.

In our analyses we obtain the allowed range of the coefficient
of each anomalous quartic operator and compare the results with
those coming from indirect measurements \cite{loops,he:yuan}, as
well as the attainable limits at future $e^+e^-$ colliders
\cite{lih,barger,boos,he,vvv}. In addition to the discovery of an
anomalous behavior of the cross section for the production of a
vector boson pair, it is important to identify the possible
source of this deviation.  Depending on the particular
operator(s) responsible for the deviations, we could have some
hint about the underlying physics that generates this departure
from the SM predictions. This can be achieved by the comparative
analysis of the different reactions since distinct operators
contribute differently to each possible two boson final states.

The paper is organized as follows. In the next Section, we
summarize the model independent formalism and present the
respective chiral Lagrangians describing the anomalous quartic
couplings among the gauge bosons. In Section III, we analyze both
the signals and backgrounds involved in the production of a
vector boson pairs accompanied by two jets. We also establish the
best cuts to improve the signal over background ratio. Our final
results for the cross sections are presented in Section IV, in
terms of the chiral Lagrangian coefficients. The final section
contains our general conclusions.

\section{Chiral Lagrangians}

When the Higgs boson is a strongly interacting particle or when
it is absent from the physical particle spectrum, one is led to
consider the most general effective Lagrangian which employs a
nonlinear representation of the spontaneously broken $SU(2)_L
\otimes U(1)_Y$ gauge symmetry \cite{Appelquist}. The resulting
chiral Lagrangian is a non-renormalizable non-linear
$\sigma$--model coupled in a gauge-invariant way to the
Yang-Mills theory.  This model independent approach incorporates
by construction the low-energy theorems \cite{cgg}, which predict
the general behavior of Goldstone boson amplitudes, irrespective
of the details of the symmetry breaking mechanism. This
low--energy effective theory should be valid up to some energy
scale smaller than $4\pi v \simeq 3$ TeV, where new physics would
come into play to avoid unitarity violation in vector--boson
scattering \cite{unit}.

In order to specify the effective Lagrangian, one must fix the
symmetry breaking pattern. We considered that the system presents
a global $SU(2)_L \otimes SU(2)_R$ symmetry that is broken to
$SU(2)$. With this choice, following the notation of Ref.\
\cite{Appelquist}, the building block of the chiral Lagrangian is
the dimensionless unimodular matrix field $\Sigma(x)$, which
transforms under $SU(2)_L \otimes SU(2)_R$ as $(2,2)$,
\begin{equation}
\Sigma(x) ~=~ \exp\left[ i \frac{\varphi^a(x) \tau^a}{v}\right] \; ,
\end{equation}
where the $\varphi^a$ fields are the would-be Goldstone fields and
$\tau^a$ ($a=1$, $2$, $3$) are the Pauli matrices.  The $SU(2)_L
\otimes U(1)_Y$ covariant derivative of $\Sigma$ is defined as
\begin{equation}
D_\mu \Sigma ~\equiv~ \partial_\mu \Sigma 
+ i g \frac{\tau^a}{2} W^a_\mu \Sigma -
i g^\prime \Sigma \frac{\tau^3}{2} B_\mu \; .
\end{equation}

The lowest-order terms in the derivative expansion of the effective
Lagrangian are
\begin{equation}
{\cal L}^{(2)} = \frac{v^2}{4} 
\hbox{Tr} \left [ \left ( D_\mu \Sigma \right )
^\dagger \left ( D^\mu \Sigma \right ) \right ]
+ \beta_1 g'^2\frac{v^2}{4} \left ( \hbox{Tr}
\left [ T V_\mu \right ] \right )^2
\; .
\label{lagran2}
\end{equation}
where we have introduced the auxiliary quantities $T \equiv
\Sigma \tau^3 \Sigma^\dagger$ and $V_\mu \equiv \left ( D_\mu
\Sigma \right ) \Sigma^\dagger$ which are $SU(2)_L$-covariant and
$U(1)_Y$-invariant. Notice that $T$ is not invariant under
$SU(2)_C$ custodial due to the presence of $\tau^3$.

The first term of the above equation is responsible for giving
mass to the gauge bosons $W^\pm$ and $Z$ for $v=(\sqrt{2}
G_F)^{-1}$. The second term violates the custodial $SU(2)_C$
symmetry and contributes to $\Delta\rho$ at the tree level, being
strongly constrained by the low-energy data. This term can be
understood as the low-energy remnant of the high-energy custodial
symmetry breaking physics, which has been integrated out above a
certain scale $\Lambda$.  Moreover, at the one-loop order, it is
also required in order to cancel the divergences in $\Delta\rho$,
arising from diagrams containing a hypercharge boson in the loop
\cite{Appelquist}.  This subtraction renders a finite
$\Delta\rho$, although dependent on the renormalization scale.

At the next order in the derivative expansion ($p^4$), there are
many operators that can be written down \cite{Appelquist}. We
shall restrict our analyses to the ones that exhibit genuine
quartic vector-boson interactions, {\em i.e.} that do not have
triple gauge--boson vertices associated to these quartic
couplings. These operators are
\begin{eqnarray}
{\cal L}^{(4)}_4 &=& \alpha_4\left[{\rm{Tr}}
\left(V_{\mu}V_{\nu}\right)\right]^2
\label{eff:4}
\;, \\
{\cal L}^{(4)}_5 &=& \alpha_5\left[{\rm{Tr}}
\left(V_{\mu}V^{\mu}\right)\right]^2
\;, \\
{\cal L}^{(4)}_6 &=& \alpha_6 \; {\rm{Tr}}\left(V_{\mu}V_{\nu}\right)
{\rm{Tr}}
\left(TV^{\mu}\right){\rm{Tr}}\left(TV^{\nu}\right) \;, \\
{\cal L}^{(4)}_7 &=& \alpha_7\;{\rm{Tr}}\left(V_{\mu}V^{\mu}\right)
\left[{\rm{Tr}}\left(TV^{\nu}\right)\right]^2
\;, \\
{\cal L}^{(4)}_{10} &=& \frac{\alpha_{10}}{2}
\left[{\rm{Tr}}\left(TV_{\mu}\right)
\;{\rm{Tr}}\left(TV_{\nu}\right)\right]^2
\; .
\label{eff:10}
\end{eqnarray}
These Lagrangian densities lead to quartic vertices involving
gauge bosons and/or Goldstone bosons. In the unitary gauge, there
are new  anomalous contributions to the $ZZZZ$ vertex coming from
all five operators, to the $W^+W^-ZZ$ vertex from all operators
except ${\cal L}^{(4)}_{10}$, and to $W^+ W^- W^+ W^-$
interaction arising from ${\cal L}^{(4)}_4$ and ${\cal
L}^{(4)}_5$.  Moreover, the interaction Lagrangians ${\cal
L}^{(4)}_6$, ${\cal L}^{(4)}_7$, and ${\cal L}^{(4)}_{10}$
violate the $SU(2)_C$ custodial symmetry.  Notice that the
quartic couplings involving photons remain untouched by the
genuinely quartic anomalous interactions at the order $p^4$. The
Feynman rules for the quartic couplings generated by these
operators can be found in the last article of Ref.\
\cite{Appelquist}.

\section{Signals and Backgrounds}

In our analyses, we study the strongly interacting electroweak
breaking sector at the LHC via the scattering of weak vector
bosons that are radiated off quarks. We considered the following
processes involving the four--gauge--boson interactions
(\ref{eff:4})--(\ref{eff:10}),
\begin{mathletters}
\label{pro}         
\begin{eqnarray}
p p &\to& W^+ W^-  \; j \; j \; ,
\label{wpwm}    \\    
p p &\to& W^- W^-  \; j \; j \; ,
\label{wmwm}  \\
p p &\to& W^+ W^+ \; j \; j \; ,
\label{wpwp}  \\
p p &\to& W^+ Z~  \; j \; j   \; ,
\label{wpz}  \\
p p &\to& W^- Z~  \; j \; j   \; ,
\label{wmz}  \\
p p &\to& Z~  Z~  \; j \; j \; .
\label{zz}
\end{eqnarray}
\end{mathletters}
We evaluated the complete set of QCD and electroweak scattering
amplitudes for the above processes, {\it i.e.\/} we did not use
neither the effective $W$ approximation \cite{evb} nor the
equivalence theorem \cite{equ}. Therefore, we were able to keep
track of the full correlation in the matrix elements, as well as
the interference between the anomalous and SM contributions.
Moreover, we took into account not only the electroweak
contributions but also the ${\cal O}(\alpha^2 \alpha_S^2)$ ones.
For the sake of clarity, we show in Table \ref{tab:coup} the
anomalous interactions that contribute to each of the reactions
(\ref{pro}). This table indicates the strategy that should be
followed to understand the origin of the possible deviations from
the SM.

The calculation of the matrix elements was performed numerically
using two distinct tools. On one hand, we evaluated the
scattering amplitudes in the unitary gauge using the HELAS
package \cite{helas}, with the SM contribution being generated by
Madgraph \cite{madg}. In this case, we wrote special subroutines
to evaluate the anomalous contributions
(\ref{eff:4})--(\ref{eff:10}) to the vector--boson
self--interactions.  On the other hand, the same processes were
evaluated using the CompHEP package \cite{comp}. The $p^4$ chiral
effective lagrangian was implemented into CompHEP in the unitary
and the 't Hooft--Feynman gauges. Despite the Feynman rules in
the 't Hooft--Feynman gauge being cumbersome, this gauge
maximizes the CompHEP performance and allows us to double check
our calculations by comparing the results in two gauges. The
results from HELAS/Madgraph and CompHEP were confronted and they
indeed agreed.

\begin{figure} 
\begin{center}
\vspace*{-3cm}
\hspace*{-2cm}
\mbox{\epsfxsize=1.1\textwidth \epsffile{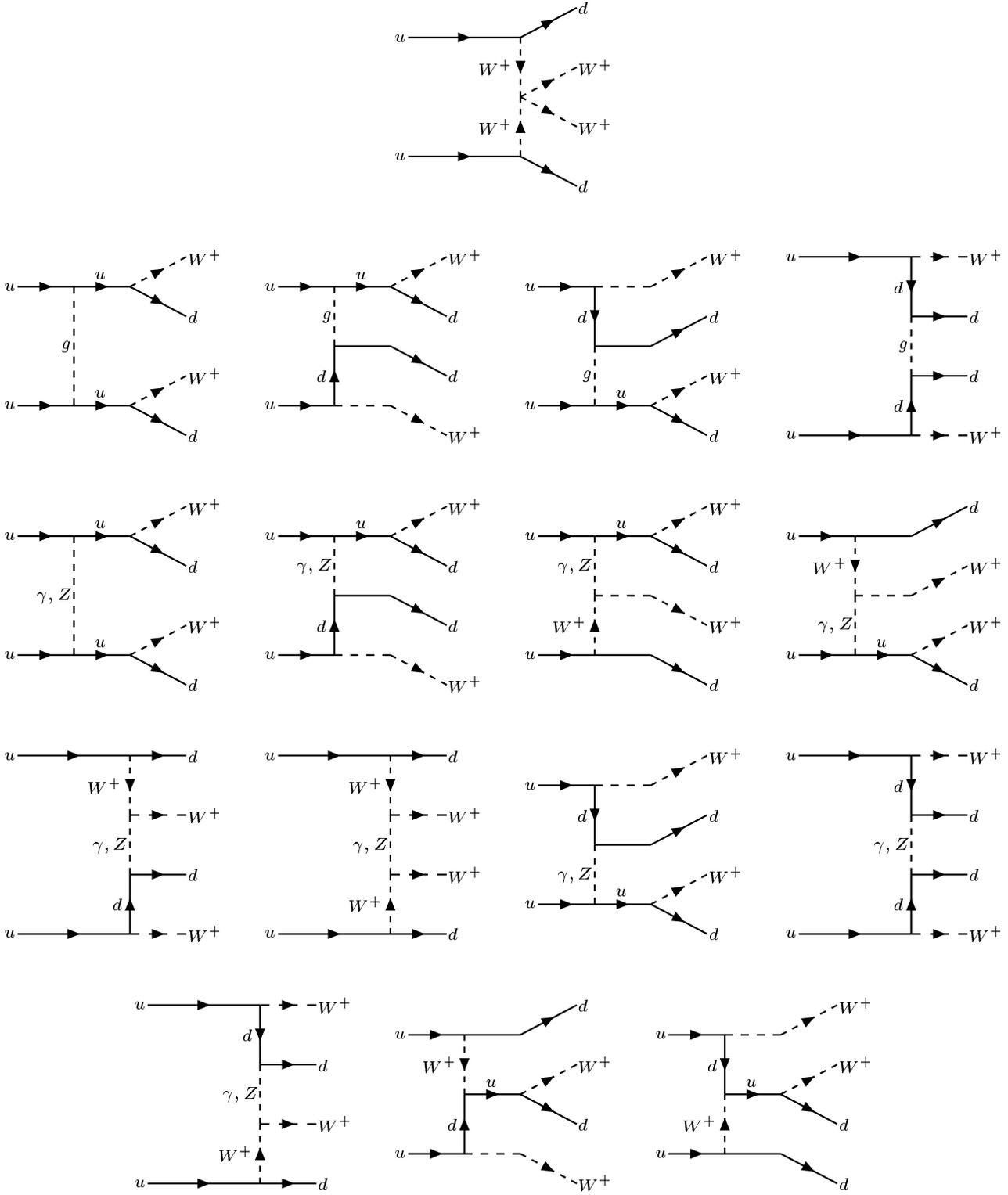}}
\vspace*{-4cm}
\end{center}
\caption{Complete set of Feynman diagrams contributing to the
process $uu \to W^+W^+ dd$.}
\label{diag}
\end{figure}

The evaluation of the processes (\ref{pro}) requires a very large
computing power. The complexity of this calculation can inferred
from the large number of diagrams involved. For instance, there
are 1918 Feynman diagrams contributing to the $W^+W^-$ final
state, while for $W^+Z$ there are 1503, and 978 for $ZZ$.  As an
illustration, we present, in Fig.~\ref{diag}, the complete set of
Feynman diagrams for the subprocess $u u \to W^+ W^+ dd$ which
contributes to the $W^+W^+$ production (\ref{wpwp}).  The first
diagram in this figure receives contributions from the anomalous
interactions, giving rise to the signal, while all other graphs
correspond to QCD and electroweak backgrounds. We neglected in
our analyses the small contribution coming from subprocesses
exhibiting two sea quarks in the initial state.

Strongly interacting symmetry breaking sectors modify the
dynamics of longitudinal vector bosons. However, it is impossible
to determine the polarization of vector bosons on an
event--by--event basis, and consequently, we have to work harder
to extract the SEWS signal. Taking into account that the
electroweak production of transversely polarized vector bosons is
approximately independent of the Higgs boson mass, and that the
$V_L V_L$ production is small for light Higgs bosons
\cite{bagger}, we define the signal for SEWS as an excess of
events in the $V V$ scattering channels with respect to the
SM model with a light Higgs, {\em i.e.}
\begin{equation}
\sigma_{\text{signal}} \equiv \sigma (\alpha_i) 
-  \sigma_{SM} \Bigr |_{M_H =100 \text{ GeV}} \;\;\; ,
\label{sig:def}
\end{equation}
where we sum over the vector-boson polarizations.  In principle,
we might have a signal even for $\alpha_i\equiv 0$ since there is
no Higgs in our model to cut off the growth of the scattering
amplitudes. In this case, we should also study whether it is
possible to establish that the anomalous couplings $\alpha_i$ are
compatible with zero or not.

In the effective--$W$ approximation \cite{evb}, the signal is
described by the scattering $V_L V_L \to V_L V_L$. This process,
however, does not respect the unitarity of the partial--wave
amplitudes $(a_\ell^I)$ at large subprocess center--of--mass
energies $M_{VV}$ \cite{boos}. Therefore, the chiral expansion is
valid only for values of $M_{VV}$ and $\alpha_i$ such that $|
a_\ell^I| \simeq 1/2$. For higher $VV$ invariant masses,
rescattering effects are important to unitarize the amplitudes.
Taking into account this fact, we conservatively restricted our
analyses to invariant masses $M_{VV}<$ 1.25 TeV, which guarantees
that the unitarity constraints are always satisfied. In the cases
where it is not possible to reconstruct the $VV$ invariant mass
from the leptonic decay products, this requirement corresponds to
a sharp--cutoff unitarization \cite{unit2}.

Since we evaluated the full matrix elements for the processes
(\ref{pro}), summed over the vector--boson polarizations, several
backgrounds were automatically included, {\it e.g.\/} the ${\cal
O}(\alpha^4)$ and ${\cal O}(\alpha^2 \alpha_S^2)$ irreducible
backgrounds $q q~ \to q q V_T V_T~ (V_L V_T)$.  In addition to
that we also evaluated the ``continuum'' $VV$ production, $q q
(gg) ~\to ~ gg~ VV$, where the vector bosons are produced in
association with gluons.  Another important background is
top--quark pair production, {\it i.e.\/} $q q (gg) ~\to ~
t\bar{t} ~ \to ~ W^+W^- b\bar{b}$ which was also taken into
account, since we considered the $W^+W^- b\bar{b}$ final state.
Moreover, triple gauge boson production also contribute to the
$VVjj$ signature when one of the three boson decays hadronically.
In principle, we should explicitly include further backgrounds
like the associated production of $t\bar{t}$ pairs accompanied by
a $W^\pm$ or a $Z$, however, these contributions are negligible
once we applied the jet veto and tag cuts described below
\cite{bagger}.

One should stress the importance of the jet--tagging and
jet--vetoing cuts since the background can be efficiently
suppressed by cutting in the jet rapidities and momenta
\cite{bagger}. In order to understand that, we must recall that
the spectra of transversely ($f^T_{W/e}$) and longitudinally
($f^L_{W/e}$) polarized $W$ in the effective $W$ approximation
are given by
\begin{eqnarray}
f^T_{W/e}(x, p_T) &=& \frac{\alpha}{4\pi \sin^2\theta_W}~ 
\frac{1+(1-x)^2}{2x}~ \nonumber \\
&& \times \frac{p_T^2}{[p_T^2 + (1-x) M_W^2]^2} \; ,
\\
f^L_{W/e}(x, p_T) &=& \frac{\alpha}{4\pi \sin^2\theta_W}~ \frac{1-x}
{x}~ \nonumber \\
&& \times \frac{(1-x) M_W^2}{[p_T^2 + (1-x) M_W^2]^2} \; ,
\end{eqnarray}
where $p_T$ is the transverse momentum of the $W^\pm$ (jet). From
the above expressions, we can learn that the transversely
polarized $W$ possess a higher $p_T$ than the longitudinally
polarized ones. Therefore the spectator jets associated with
$W^\pm_L$ are produced at large rapidities since their energies
are of the order of TeV.

Forward (backward) jets are characteristic configurations of the
signal. At the same time, jets coming from the signal are well
separated and their $p_T$ distribution does not peak near zero
because of massive vector--boson propagators. On the other hand,
the situation is opposite for some backgrounds: either their
$p_T$ distributions peak at small values due to photon, gluon or
light quark $t$--channel exchange, or they have the tendency to
be close to each other since the jets originate from gluon or
photon splitting. This remarkable difference between the signal
and the backgrounds allows us to substantially reduce the latter
by requiring the tagging of forward jets. We can further reduce
backgrounds, like $t\bar{t}$ and $VVV$, by vetoing large jet
activity in the central region of the detector \cite{tagveto}.

\begin{figure} 
\begin{center}
\begin{picture}(600,585)(0,0)
\put(0,310){\hspace*{-1.5cm}\epsfig{file=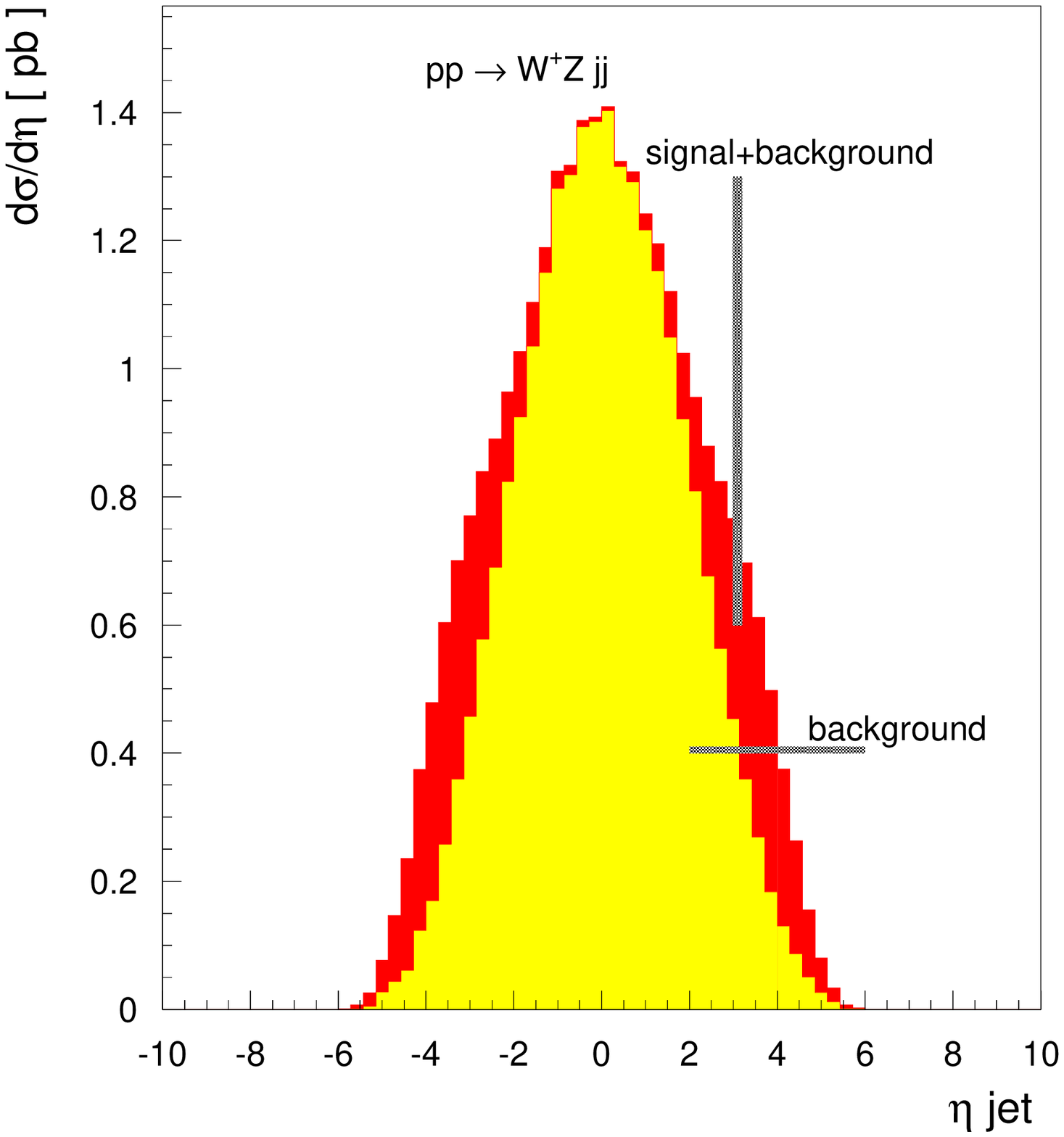,width=0.6\textwidth}
            \hspace*{-0.5cm}\epsfig{file=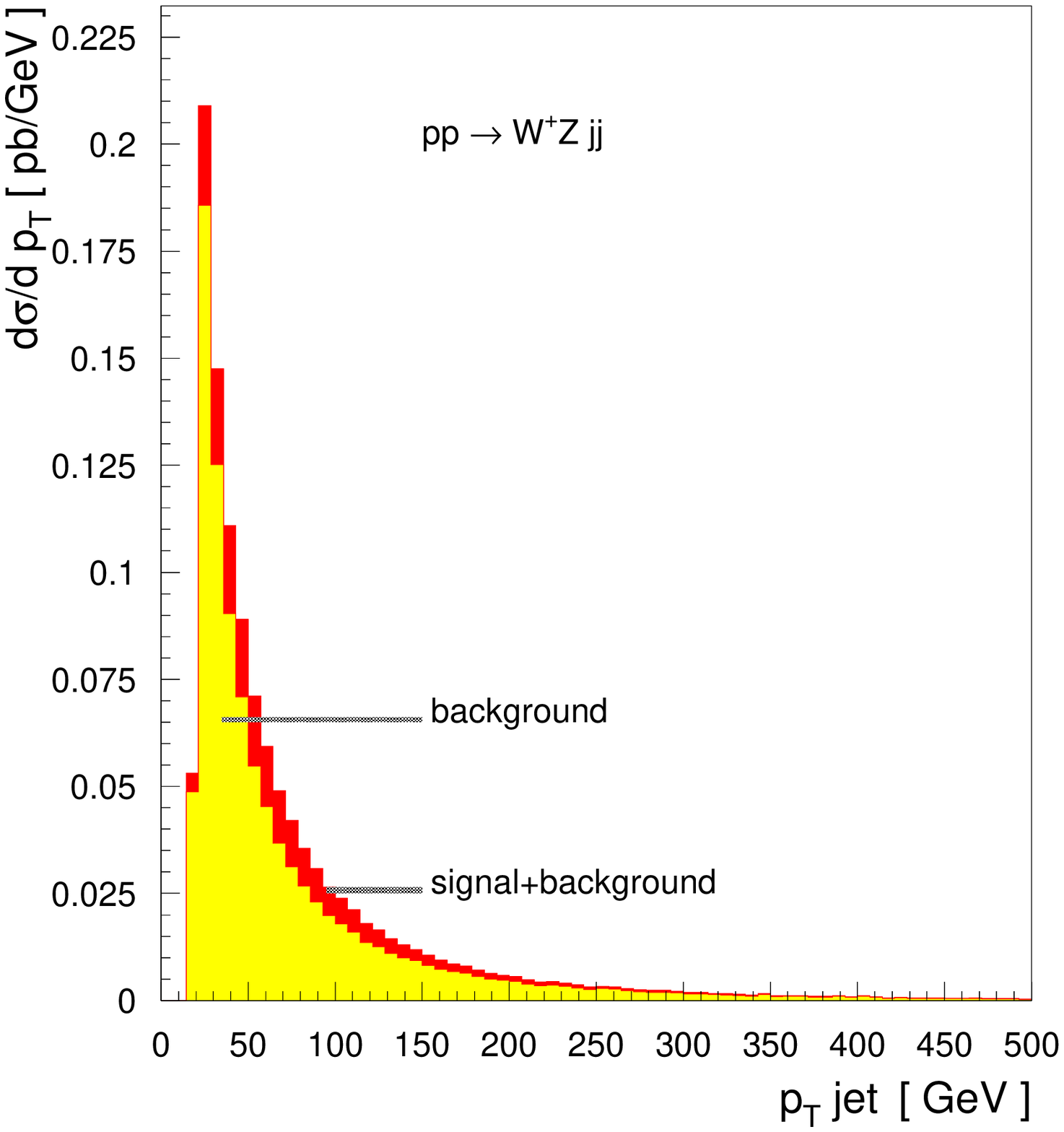,width=0.6\textwidth}}
\put(0,0){\hspace*{-1.5cm}\epsfig{file=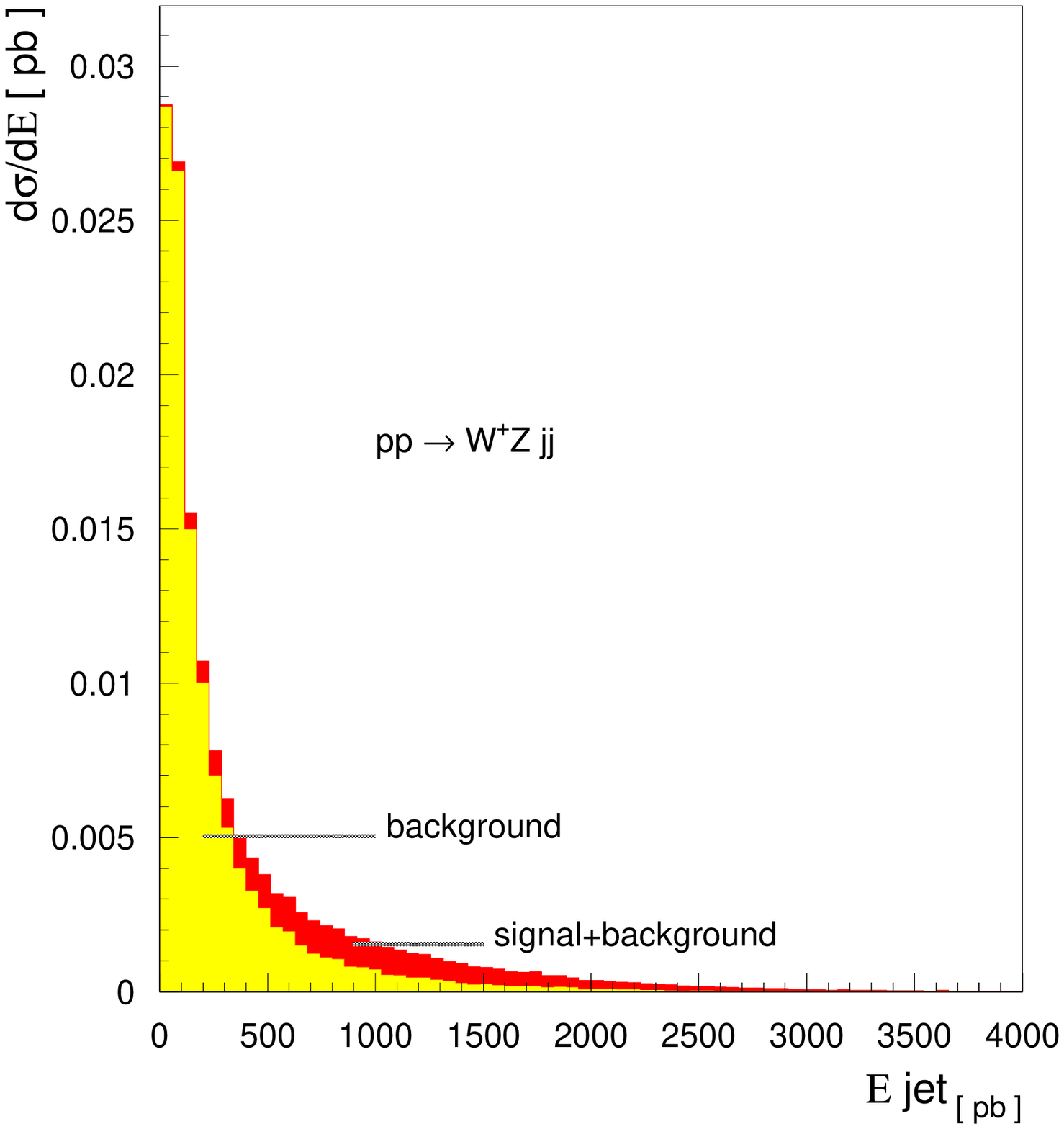,width=0.6\textwidth}
           \hspace*{-0.5cm}\epsfig{file=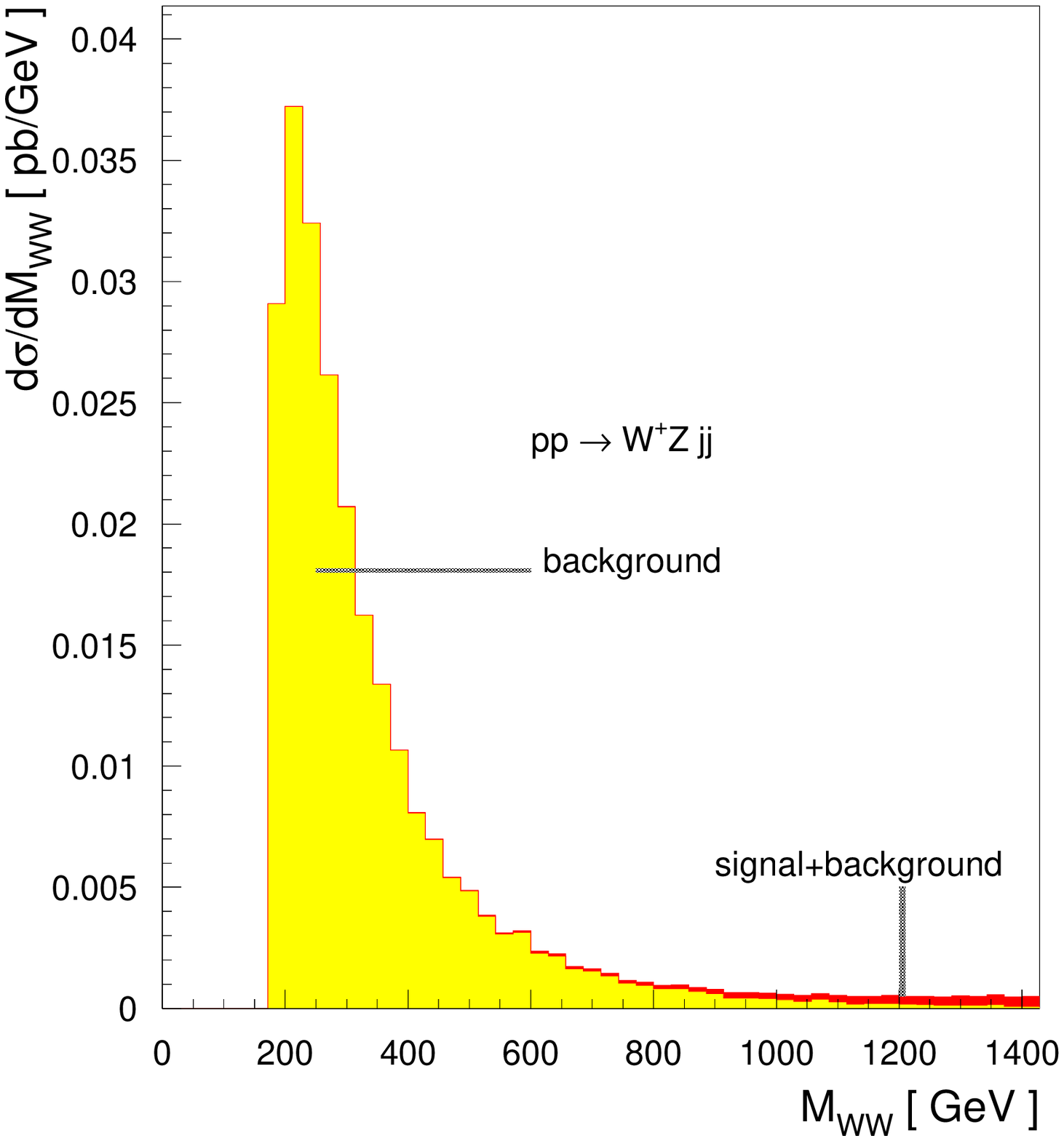,width=0.6\textwidth}}
         \put(20,  590){(a)}
         \put(325, 590){(b)} 
         \put(20,  300){(c)}
         \put(315, 300){(d)}
\end{picture}
\end{center}
\caption{Kinematical distributions for the process $p p \to W^+ Z
j j$: (a) pseudo--rapidity of the jets ($\eta_j$); (b) transverse
momentum of the jets ($p_{T_j}$); (c) energy of the jets ($E_j$);
and (d) invariant mass of the $W^+ Z$ pair ($M_{WW}$). The light
gray area stands for the background while the dark area
represents the background plus the signal associated to $\alpha_4
= 0.03$. We required that $p_T^{\text{jet}}>20$ GeV and the jet
separation $\Delta R_{jj}>0.5$.}
\label{fig:dis}
\end{figure}

In Fig.\ \ref{fig:dis}, we show some kinematical distributions
for the process $pp \to W^+ Z j j $ (\ref{wpz}). Fig.\
\ref{fig:dis}a contains the pseudo--rapidity distribution of the
jets, while we exhibit the $p_T$ (energy) distribution of the
jets in Fig.\ \ref{fig:dis}b (c), and the invariant mass
distribution of $W^+ Z$ pairs in Fig.\ \ref{fig:dis}d.  From
these figures we can see that the jets associated with the signal
are produced at large rapidities and carry a larger amount of
energy, illustrating the importance jet--tagging and jet--vetoing
cuts.

In order to suppress the backgrounds and enhance the signal for
anomalous quartic interactions we studied several kinematical
distributions for the processes (\ref{pro}), applying different
cuts on the final state particles. Our results indicate that the
cuts presented in Ref.\ \cite{bagger} are able to improve
considerably the signal/background ratio. We applied the
following set of kinematical cuts,   keeping those from the above
mentioned paper and also suggesting some additional ones that
could allow further suppression of the backgrounds:

\begin{itemize}
  
\item [$(i)$] We required the existence of two jets satisfying
$p_T > 20$ GeV, $|\eta| < 5$, and $\Delta R \equiv \sqrt{(\Delta
\eta^2 + \Delta \phi^2)} > 0.5$.  The cut in $p_T$ is important
not only to guarantee that the jets will be well defined, but
also to suppress the background due to the photon and gluon
exchanges in $t$ channel.  At the same time, the $\Delta R$ cut
is necessary, combined with the $p_T$ one, to remove the
singularity coming from gluon splitting in some background
subprocesses.

\item [$(ii)$] We applied the jet--tagging and jet--vetoing cuts
suggested by Bagger {\it et al.} (\cite{bagger}), {\it i.e.},
\begin{eqnarray*} 
&&E(j_{\text{tag}}) > 0.8 \; \text{TeV} \;\; 
(\text{except for} \; W^\pm W^\pm ) \; , \\ 
&&3.0 < \vert y(j_{\text{tag}}) \vert < 5.0 \;\;  , \; \; \; 
p_T(j_{\text{tag}}) > 40 \; \text{GeV}\; , \\
&&p_T(j_{\text{veto}}) > 60 \; \text{GeV} \;\;
(30 \; \text{GeV} \; \text{for} \; W^+W^-) \;\;  , \\ 
&& \vert  y(j_{\text{veto}}) \vert  < 3.0 \; . 
\end{eqnarray*} 
$t\bar{t}$ production gives rise to a quite large background to
the $W^+W^-jj$ signal, and consequently, the requirement of a
more stringent $p_T(j_{\text{veto}})$ cut for this process is
important to improve the signal/background ratio.

\item [$(iii)$] We also required the invariant mass of the vector
boson pair to be in the range $0.5 < M_{VV} < 1.25$ TeV. The
upper limit of this cut is quite important since it prevents the
effective operators (\ref{eff:4})--(\ref{eff:10}) to be used in a
energy regime where unitarity is violated and rescattering
effects become important. The lower limit of this cut aims to
reduce the background (see Fig.\ \ref{fig:dis}).

\end{itemize}

In this work we considered the ``gold--plated'' events where the
$W$'s and $Z$'s decay into electrons or muons, ignoring final
states associated with the hadronic decay of the vector bosons.
In order to make a more realistic estimation of the limits that
can be imposed on the anomalous parameters, one should take into
account the detection efficiency of the final state leptons. This
problem was studied in Ref.\ \cite{lhc} for $W^\pm$ and $Z$
decays in Higgs production processes. Imposing that the leptons
satisfy the following cuts
\[
|\eta^\ell| < 2 \;\; , \;\;\;  p_T^\ell > 100 \; \text{GeV} \; ,
\;\; \text{and} \;\;\;  p_T^{\text{miss}} > 100 \text{GeV},
\]
the detection efficiency for leptons originating from $W$ ($Z$) decays
is 43\% (52\%) \cite{lhc}.  We also took into account the branching
ratios of $W^\pm$ and $Z$ into electrons or muons ($\ell = e$ or
$\mu$),
\begin{eqnarray*}
&&\text{BR}(WW \rightarrow \ell \bar\nu_\ell \; \bar\ell \nu_\ell)=4.7\%
\; , \\
&&\text{BR}(W^+Z \rightarrow \ell \bar\nu_\ell \; \ell \bar\ell )=1.5\%
\; , \\
&&\text{BR}(ZZ \rightarrow 4\ell) = 0.45\% \; .
\end{eqnarray*}


\section{Results}

The most general expression for the total cross sections of the
processes (\ref{pro}) can be written as
\begin{equation}
\sigma = C_0 + \sum_j \alpha_j \; C_j +
\sum_{j \le k} \alpha_j \; \alpha_k \; C_{j-k} \; ,
\label{sig:gen}
\end{equation}
where $j,k=4$, $5$, $6$, $7$, or $10$ and $C_0$ is the cross
section for $\alpha_j \equiv 0$. In our calculations, we applied
the cuts $(i) - (iii)$ and used the CTEQ3M parton distributions
\cite{cteq3}, with $Q^2$ equal to the invariant mass of the parton
system. We present in Table \ref{tab:sig} our results for the
coefficients $C_0, C_j, C_{j-k}$, as well as for the SM with a Higgs of
mass $M_H=100$ GeV ($C_{\text{SM}}$)

\begin{figure} 
\begin{center}
\mbox{\epsfig{file=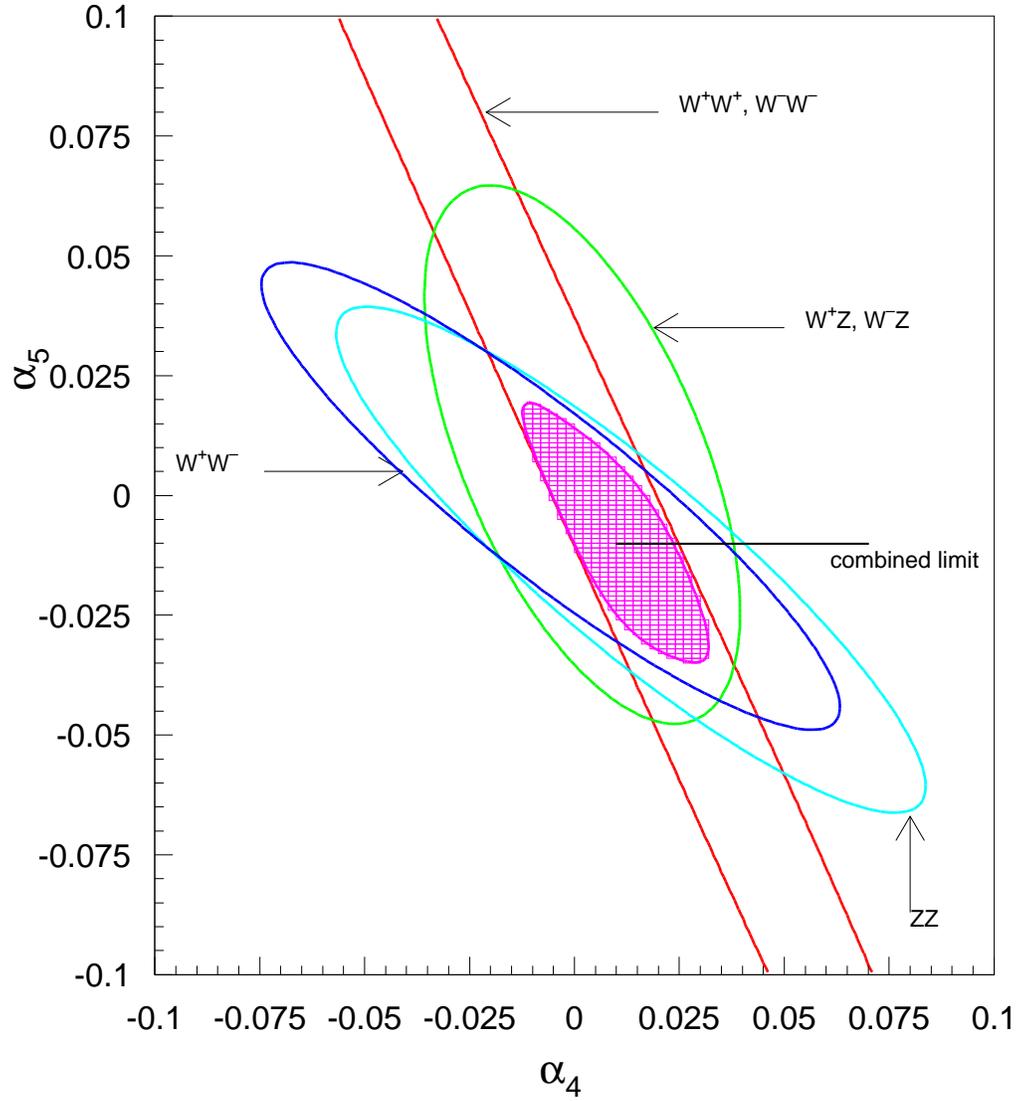,width=0.8\textwidth}}
\end{center}
\caption{90\% CL exclusion region in the $\alpha_4 \times
  \alpha_5$ plane for the $W^+W^-$, $W^\pm W^\pm$, $W^\pm Z$, and $ZZ$
  channels. We applied all cuts and efficiencies discussed in the text and
  assumed that all $SU(2)_C$ violating couplings vanish and an integrated
  luminosity of 100 fb$^{-1}$.}
\label{fig:45}
\end{figure}

Given our definition of the signal (\ref{sig:def}) and the above
parametrization of the anomalous cross section we can easily
obtain the LHC attainable limits on any combination of genuinely
quartic anomalous couplings. We exhibit in Fig.\ \ref{fig:45} the
90\% CL exclusion region in the plane $\alpha_4 \times \alpha_5$
for each process (\ref{pro}) independently, assuming an
integrated luminosity ${\cal L} = $ 100 fb$^{-1}$ and taking
properly into account the detection efficiencies and leptonic
branching ratios. In this analysis, we assumed that the $SU(2)_C$
violating interactions vanish. As we can see, the $W^\pm Z$,
$ZZ$, and $W^+W^-$ productions lead to similar bounds while the
$W^\pm W^\pm$ give rise to somewhat weaker limits. Combining all
channels allow us to improve the limits by a factor of
approximately 2

\begin{figure} 
\begin{center}
\mbox{\epsfig{file=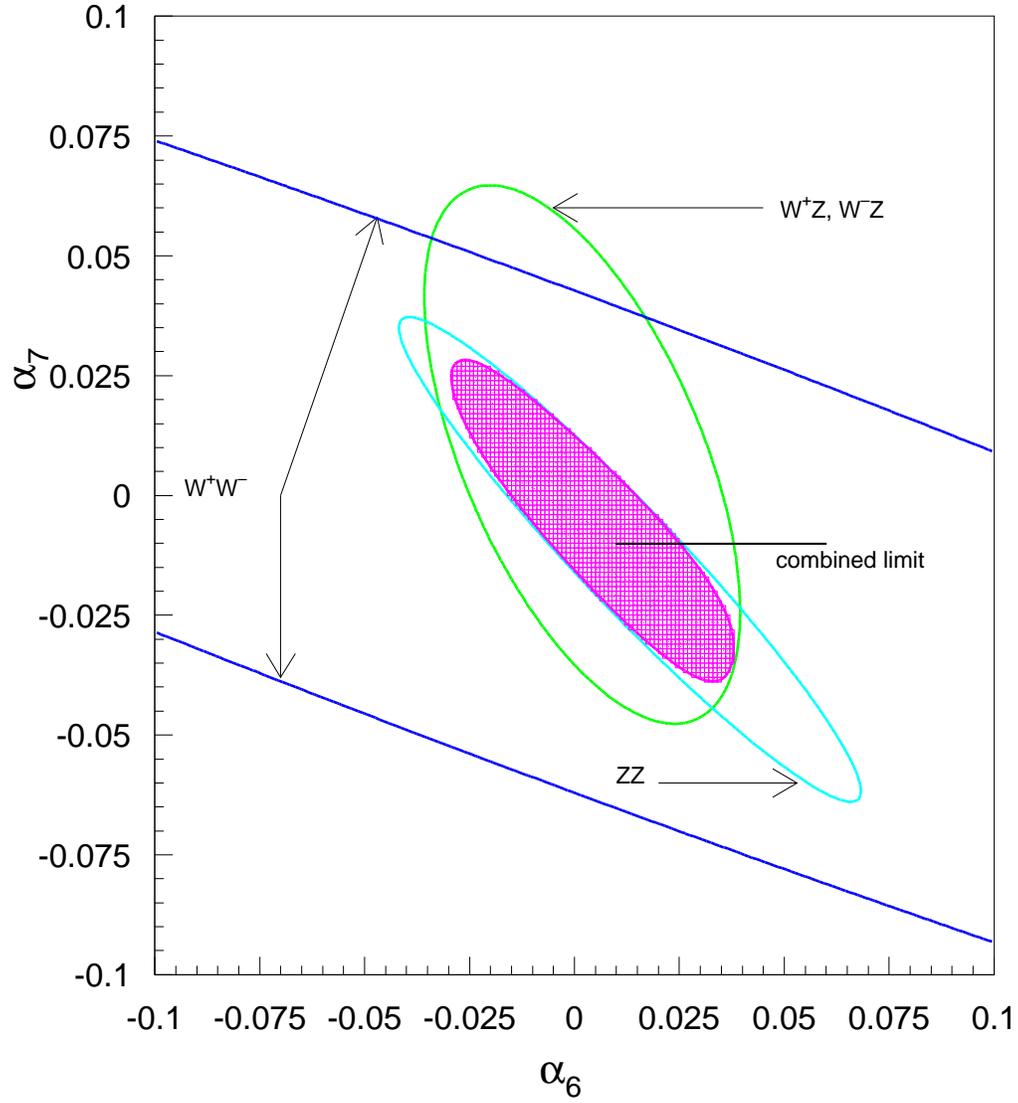,width=0.8\textwidth}}
\end{center}
\caption{90\% CL exclusion region in the $\alpha_6 \times
  \alpha_7$ plane from the $W^+W^-$, $W^\pm Z$, and $ZZ$ productions.  We
  applied all cuts and efficiencies discussed in the text and assumed that
  $\alpha_4 = \alpha_5 = \alpha_{10} = 0$ and an integrated luminosity of 100
  fb$^{-1}$.}
\label{fig:67}
\end{figure}

Fig.\ \ref{fig:67} contains the 90\% CL exclusion region in the
$\alpha_6 \times \alpha_7$ plane for $\alpha_4 = \alpha_5 =
\alpha_{10} =0$ and an integrated luminosity of 100 fb$^{-1}$.
The $W^\pm W^\pm$ production does not give rise to any bound
since these interactions possess only $ZZZZ$ and $W^+W^-ZZ$
anomalous couplings.  Moreover, the production of $W^+W^-$ pairs
leads to weak bounds since these couplings contribute to this
final state only through the subprocess $ZZ\to W^+W^-$, which is
suppressed.  The best limits come from the $ZZ$ pair production
and the combined limits of $ZZ$ and $W^\pm Z$ productions are
only slightly better than the $ZZ$ bounds.

\begin{figure}
\begin{center}
\mbox{\epsfig{file=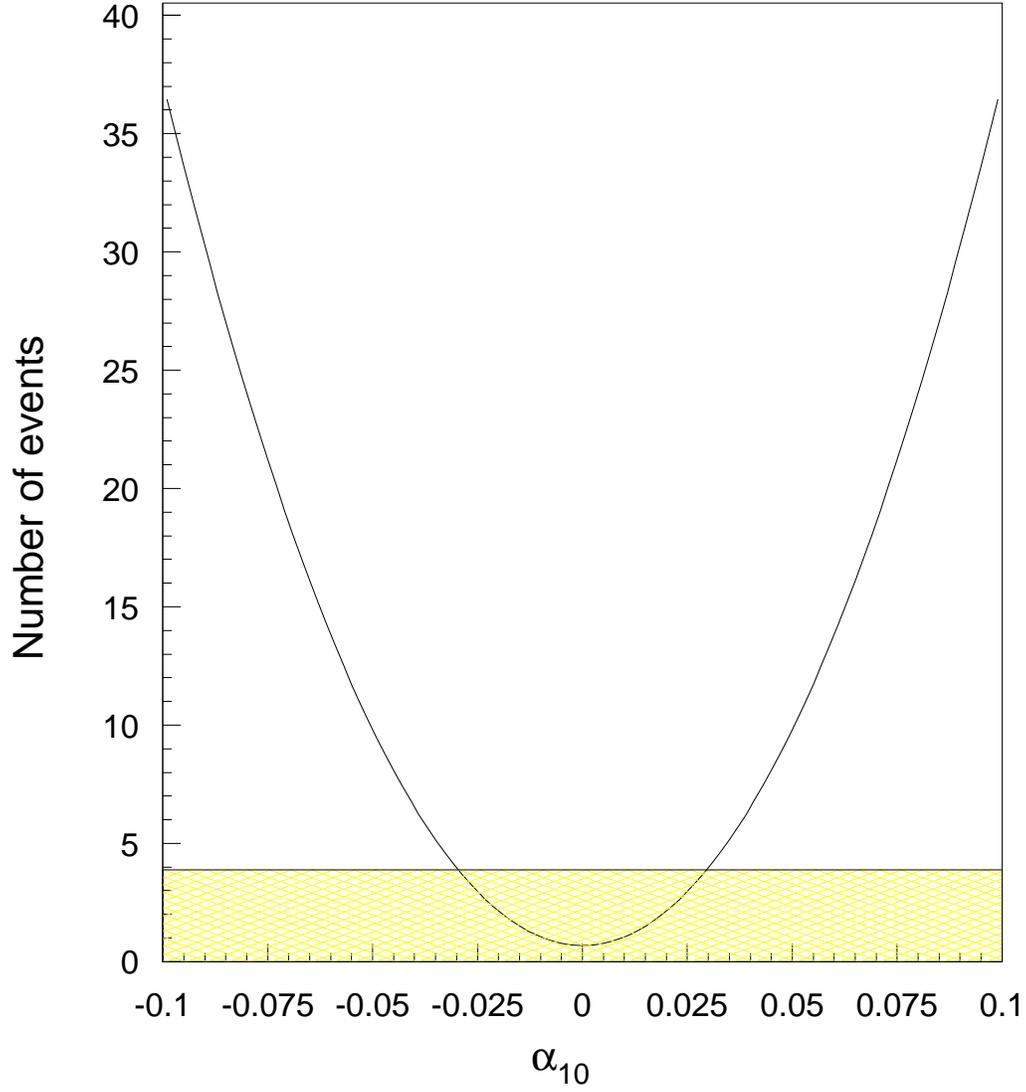,width=0.8\textwidth}}
\end{center}
\caption{Number of events for $ZZ$ production as a function of
  $\alpha_{10}$ where the horizontal line represents a 90\% CL effect.  We
  applied all cuts and efficiencies discussed in the text and assumed that
  $\alpha_4 = \alpha_5 = \alpha_6 = \alpha_7 =0 $ and an integrated luminosity
  of 100 fb$^{-1}$.}
\label{fig:10}
\end{figure}

The anomalous interaction $\alpha_{10}$ modifies only the $ZZjj$
production since it alters solely the vertex $ZZZZ$. We present
in Fig.\ \ref{fig:10} the limits that can be obtained on this
coupling from the $ZZ$ pair production for $\alpha_4 = \alpha_5 =
\alpha_6 = \alpha_7 = 0 $ and an integrated luminosity of 100
fb$^{-1}$. Therefore, this coupling is the one that will be less
constrained at the LHC.

Table III shows the limits on each coupling $\alpha_i$, $i =$4,
5, 6, 7, and 10, taking into account our results presented in
Figs.\  \ref{fig:45}, \ref{fig:67}, and \ref{fig:10}.  These
limits were obtained under the assumption that only one anomalous
parameter is non--vanishing. For the sake of comparison, this
Table also contains the present indirect bounds on these
anomalous couplings obtained from the precision measurements at
the $Z$ pole \cite{loops} for a scale of new physics $\Lambda=2$
TeV. As we can see, the direct bounds on $\alpha_4$ and
$\alpha_5$ that can be obtained from the $V V j j$ production at
LHC are more restrictive than the present limits by one order of
magnitude in some cases. Nevertheless, the attainable direct
limits on the $SU(2)_C$ violating interactions $\alpha_6$,
$\alpha_7$, and $\alpha_{10}$ are of the same order of the
present indirect limits.

It is also important to devise a strategy to disentangle the
anomalous couplings in case a departure from the SM prediction is
observed.  In fact the simultaneous analysis of the $W^\pm W^\pm
jj$, $W^+W^-jj$, $W^\pm Zjj$, and $ZZjj$ productions allows us to
narrow down the anomalous couplings associated to the observed
effect. The anomalous couplings $\alpha_4$ and $\alpha_5$ possess
the distinctive characteristic of giving rise to observable
effects for all processes $VVjj$. On the other hand, the
couplings $\alpha_6$ and $\alpha_7$ lead to large signals in the
channels $W^\pm Z$ and $ZZ$ without any excess in the $W^\pm
W^\pm$ reaction. Finally the anomalous coupling $\alpha_{10}$
gives rise only to an excess of events in the $ZZ$ channel.  The
effects of $\alpha_4$ and $\alpha_5$ ($\alpha_6$ and $\alpha_7$)
can only be separated if we have additional information like the
triple gauge--boson production at the NLC, where the
$\alpha$'s appears in different combinations for the different
channels.

\section{Conclusions}

In this paper we presented the first complete calculation of the
reaction $pp \to VV jj$ taking into account anomalous quartic
vector--boson couplings.  Our calculations were done at tree
level in two different gauges and without any approximation, such
as the effective $W$ one or the equivalence theorem.  Our results
show the ability of the LHC to shed some light on the electroweak
symmetry breaking sector and to look for a possible signal of
strongly interacting electroweak symmetry breaking.

The attainable LHC limits for the quartic anomalous parameters
are tighter than the present indirect bounds
\cite{loops,he:yuan}, improving them by one order of magnitude in
some cases. The LHC bounds are also one order of magnitude better
than those which could be obtained from the study of triple
gauge--boson production at the Next Linear Collider (NLC)
\cite{lih,he,vvv}. Notwithstanding, the study of the reaction
$VVjj$ at the NLC running at TeV energies \cite{boos,he} will be
able to improve the LHC limits by a factor of 2 to 8, depending
on the specific couplings.

In our analyses, we assumed that the detection efficiencies of
electrons and muons are the ones obtained from the production of
heavy Higgs bosons. For a more realistic study one should 
construct a complete Monte Carlo generator including the
vector--boson decays and detector resolution \cite{future}. Such
generator will allow not only to improve the leptonic cuts but
also to study the hadronic decay channels of one of the gauge
bosons, which could improve the limits on the anomalous
couplings. We believe that even assuming this more realistic
situation, the bounds presented in this paper will not change
significantly.

\acknowledgments

One of the authors (A.S.B.) is grateful to A.\ Pukhov for
important improvements of the CompHEP package and to A.\ Solomin
for the help in computing facilities.  This work was supported by
Funda\c{c}\~ao de Amparo \`a Pesquisa do Estado de S\~ao Paulo
(FAPESP), by Conselho Nacional de Desenvolvimento Cient\'{\i}fico
e Tecnol\'ogico (CNPq), by DGICYT under grant PB95-1077, and by
CICYT under grant AEN96--1718.



\newpage

\begin{table}
\begin{tabular}{||c||c||c||c||}
Coupling & $W^\pm W^\pm$ & $W^\pm Z$ & $ZZ$ \\
\hline 
\hline
$\alpha_{4,5}$ & Yes & Yes & Yes \\
\hline
$\alpha_{6,7}$ & No  & Yes & Yes  \\
\hline
$\alpha_{10}$ & No  & No  & Yes  \\
\end{tabular}
\medskip\medskip
\caption{The processes affected by the different quartic couplings 
(\protect\ref{eff:4})--(\protect\ref{eff:10}).}
\label{tab:coup}
\end{table}


\begin{table}
\begin{tabular}{||c||c||c||c||c||c||c||}
               &$W^+W^-$&$W^+W^+$ &$W^-W^-$ &$W^+Z$  &$W^-Z$   & $ZZ$   \\
\hline 
\hline
$C_{\text{SM}}$
                & 0.049 &0.0044   &0.0009   &0.018    &0.0070   & 0.0044 \\
\hline
$C_{0}$         & 0.050 &0.0061   &0.0011   &0.019    &0.0074   & 0.0056 \\
\hline
$C_{4}$         & 0.21  & $-$0.38 &$-$0.062 &$-$0.14  &$-$0.062 & 0.066 \\
\hline
$C_{5}$         & 0.27  & $-$0.19 &$-$0.034 &$-$0.12  &$-$0.057 & 0.20  \\
\hline
$C_{6}$         & 0.036 & ---     & ---     &$-$0.14  &$-$0.062 & 0.066 \\
\hline
$C_{7}$         & 0.11  & ---     & ---     &$-$0.12  &$-$0.057 & 0.20  \\
\hline
$C_{10}$        & ---   & ---     & ---     & ---     & ---     & $-$0.00012\\
\hline
$C_{4-4}$       & 18.   & 27.     & 4.3     &14.      & 5.4     & 13.   \\
\hline
$C_{5-5}$       & 36.   & 7.2     & 1.2     &6.3      & 2.4     & 23.   \\
\hline
$C_{6-6}$       & 0.67  & ---     & ---     &14.      & 5.4     & 49.   \\
\hline
$C_{7-7}$       & 5.7   & ---     & ---     &6.3      & 2.4     & 58.   \\
\hline
$C_{10-10}$     & ---   & ---     & ---     & ---     & ---     & 47.   \\
\hline
$C_{4-5}$       & 46.   & 28.     & 4.4     & 11.     & 4.2     & 31.   \\
\hline
$C_{4-6}$       & 1.4   & ---     & ---     & 29.     & 11.     & 50.   \\
\hline  
$C_{4-7}$       & 3.6   & ---     & ---     & 11.     & 4.2     & 55.   \\
\hline
$C_{4-10}$      & ---   & ---     & ---     & ---     & ---     & 47.   \\
\hline
$C_{5-6}$       & 4.0   & ---     & ---     & 11.     & 4.2     & 54.   \\
\hline
$C_{5-7}$       & 12.   & ---     & ---     & 13.     & 4.8     & 69.   \\
\hline
$C_{5-10}$      & ---   & ---     & ---     & ---     & ---     & 47.   \\
\hline
$C_{6-7}$       & 3.7   & ---     & ---     & 11.     & 4.2     & 102.  \\
\hline
$C_{6-10}$      & ---   & ---     & ---     & ---     & ---     & 94.   \\
\hline
$C_{7-10}$      & ---   & ---     & ---     & ---     & ---     & 94.   \\
\end{tabular}
\medskip\medskip
\caption{Coefficients of the different combinations of coupling
constants contributing to the total cross section in pb [see Eq.\
(\protect\ref{sig:gen})], and also for the SM with a light Higgs ($M_H
= 100$ GeV). These results were obtained applying the cuts $(i)-(iii)$.
}
\label{tab:sig}
\end{table}


\begin{table}
\begin{tabular}{|| c || c || c ||}
&& \\
Coupling & Indirect Limits 	& LHC Limits \\
         & ($\times 10^{-3}$)  	& ($\times 10^{-3}$) \\
&& \\
\hline
\hline
$\alpha_4$ &$-160. \leq \alpha_4 \leq 54. $
           &$-3.5  \leq \alpha_4 \leq 15. $
\\ \hline
$\alpha_5$ &$-410. \leq \alpha_5 \leq 13. $
           &$-7.2  \leq \alpha_5 \leq 13. $
\\ \hline
$\alpha_6$ &$-27. \leq \alpha_6 \leq 8.9 $
           &$-13. \leq \alpha_6 \leq 13. $
\\ \hline
$\alpha_7$ &$-26. \leq \alpha_7 \leq 8.5 $
           &$-13. \leq \alpha_7 \leq 11. $
\\ \hline
$\alpha_{10}$&$-28. \leq \alpha_{10} \leq 9. $
             &$-29. \leq \alpha_{10} \leq 29.$ 
\end{tabular}
\label{tab:limits}
\medskip\medskip
\caption{Limits on the anomalous quartic couplings $\alpha_i$
which will be accessible at LHC, as well as the present indirect
bounds from Ref.\ \protect\cite{loops}.}
\end{table}


\onecolumn

\end{document}